\documentclass[12pt]{iopart}
\usepackage{iopams} 
\usepackage{subfigure}
\usepackage{color} 
\usepackage{graphicx}
\usepackage{verbatim} 
\newcommand*{\ra}{{\rightarrow}}

\newcommand*{\ang}[1]{\left\langle #1 \right\rangle}

\newcommand*{\pp}{{{\partial}}}

\def\Om{\Omega}
\def\Dd{\Delta}
\def\aa{\alpha}

\def\k{k}
\begin{document}

\title[Spectral scaling of the Leray-$\alpha$ model for two-dimensional turbulence]
{Spectral scaling of the Leray-$\alpha$ model for two-dimensional turbulence}

\author{Evelyn Lunasin$^1$, Susan Kurien$^2$, and Edriss S. Titi$^{3,4}$}

\address{$^1$ Department of Mathematics, University of California, San Diego (UCSD), La Jolla, CA 92093, USA\\
$^2$ Theoretical Division, Los Alamos National Laboratory, Los Alamos, NM 87545, USA\\
$^3$ Department of Mathematics and Department of Mechanical and Aerospace Engineering, University of California, Irvine (UCI), Irvine, CA 92697, USA\\  
$^4$ Department of Computer Science and Applied Mathematics, Weizmann Institute of Science, Rehovot 76100, Israel}
\ead{elunasin@math.ucsd.edu, skurien@lanl.gov, etiti@math.uci.edu}

\begin{abstract}
  We present data from high-resolution numerical simulations of the
  Navier-Stokes-$\alpha$ and the Leray-$\alpha$ models for
  two-dimensional turbulence. It was shown previously (Lunasin et al., J.
  Turbulence, 8, (2007), 751-778), that for wavenumbers $k$ such that
  $k\alpha\gg 1$, the energy spectrum of the smoothed velocity field
  for the two-dimensional Navier-Stokes-$\alpha$ (NS-$\alpha$) model
  scales as $k^{-7}$. This result is in agreement with the
  scaling deduced by dimensional analysis of the flux of the
  conserved enstrophy using its characteristic time scale. We therefore
  hypothesize that the spectral scaling of any $\alpha$-model in the
  sub-$\alpha$ spatial scales must depend only on the characteristic time scale
  and dynamics of the dominant cascading quantity in that regime of
  scales. The data presented here, from simulations of the two-dimensional Leray-$\alpha$ model, 
  confirm our hypothesis. We show that for $k\alpha\gg 1$, the energy
  spectrum for the two-dimensional Leray-$\alpha$ scales as $k^{-5}$, as expected by the characteristic time scale
  for the flux of the conserved enstrophy of the Leray-$\alpha$ model.
  These results lead to our conclusion that the dominant directly
  cascading quantity of the model equations must determine the scaling of
  the energy spectrum.
\end{abstract}
\maketitle

\section{Introduction}

In \cite{LKTT} we observed that the scaling exponent of the energy
spectrum of the two-dimensional (2d) Navier-Stokes-$\alpha$ model (NS-$\alpha$), for
wavenumbers $k$ such that $k\alpha\gg 1$, is $k^{-7}$. {\it A
  posteriori}, we saw that this scaling corresponding to that
predicted by assuming that the dynamics for $k\alpha\gg 1$ was
governed by the characteristic time scale for flux of the conserved
enstrophy. We were therefore led to speculate that (in general)
the unknown scaling exponent for any $\alpha$-model may be predicted by
the dynamical time scales for the dominant conserved quantity for that
model in the regime $k\alpha\gg 1$. In this paper, we present new
numerical simulations of the 2d Leray-$\alpha$ model which support this
hypothesis.

We measure the scaling of the energy spectra from simulations of 
two-dimensional flow, performed at a resolution of $4096^2$, 
in the limit as $\alpha \rightarrow \infty$, for
two models: the NS-$\alpha$ model \cite{FHTM,FHTP,HMR,MM} 
\begin{eqnarray}\label{nsa}
  &\pp_t \tilde{v} - \nu\Dd \tilde{v} - \tilde{u}\times\nabla\times \tilde{v} = -\nabla \tilde{p} + f,\\
&\nabla \cdot \tilde{u} =  \nabla \cdot \tilde{v} = 0,\\
&\tilde{v} = (I-\aa^2\Dd)\tilde{u}, \label{rel-n}
\end{eqnarray}
and the Leray-$\alpha$ model \cite{CHOT}
\begin{eqnarray}\label{leray-a}
  &\pp_t v - \nu\Dd v + (u\cdot\nabla) v = -\nabla p + f,\\&\nabla \cdot u =  \nabla \cdot v = 0,\\ 
  &v = (I-\aa^2\Dd)u, \label{rel-l}
\end{eqnarray}
where $v$, $u$ and $p$ are the unsmoothed velocity, smoothed velocity and the pressure respectively for the Leray-$\alpha$ model and we use ~$\tilde{}$~ to distinguish the variables in the NS-$\alpha$ model; $\nu$ is the viscosity, and $f$ is the body force. Notice that the
two systems above reduce to Navier-Stokes equations (NSE) when
$\alpha=0$.  One can think of the parameter $\alpha$ as the length
scale associated with the width of the filter which smooths $v$ (or
$\tilde{v}$) to obtain $u$ (or $\tilde{u}$).  The filter is associated
with the Green's function (Bessel potential) of the Helmholtz operator
$(I-\alpha^2\Delta)$.  We supplement both of the systems above with
periodic boundary conditions in a basic box $[0,L]^2$.

The inviscid and unforced version of the three-dimensional (3d) NS-$\alpha$ 
 was introduced in \cite{HMR} based on the Hamilton
variational principle subject to the incompressibility constraint
$\mbox{div}\; \tilde{v} =0$. 
By adding the viscous term $-\nu\Delta \tilde{v}$ and the forcing $f$ in an {\it ad hoc} fashion, the authors in \cite{CH98,CH99,CH00} and \cite{FHTM}
obtain the NS-$\alpha$ system which they named, at the time, the
viscous Camassa-Holm equations (VCHE), also known as the Lagrangian
averaged Navier-Stokes-$\alpha$ model (LANS-$\alpha$).  In references~\cite{CH98, CH99, CH00} it was found that
the analytical steady state solutions for the 3d NS-$\alpha$ model
compared well with averaged experimental data from turbulent flows in
channels and pipes for wide range of large Reynolds numbers. It was this
fact which led the authors of \cite{CH98, CH99, CH00} to suggest that
the NS-$\alpha$ model be used as a closure model for the Reynolds
averaged equations.  Since then, it has been found that there is
in fact a whole family of `$\alpha$'- models which provide similar
successful comparison with empirical data -- among these are the
Clark-$\alpha$ model \cite{CHTi, Clark}, the Leray-$\alpha$ model
\cite{CHOT}, the modified Leray-$\alpha$ model \cite{ILT} and the
simplified Bardina model \cite{CLT, LL} (see also \cite{OlTi} for a
family of similar models). 

The 3d NS-$\alpha$ model was tested numerically in \cite{CH99} and \cite{CH01}, for
moderate Reynolds number in a simulation of size $256^3$, with
periodic boundary conditions. It was observed that the large scale
features of a turbulent flow were indeed captured and there was a
roll over of the energy spectrum from $k^{-5/3}$ for $k\alpha \ll 1$ to
something steeper for $k\alpha \gg 1$, although the scaling ranges
were insufficient to enable extraction of the power law unambiguously. Other numerical
tests of the NS-$\alpha$ model were performed in
\cite{GH}, \cite{GH2}, and \cite{MM}, with similar results.

In the limit as $\alpha\rightarrow\infty$, we call the two equations
NS-$\infty$ \cite{IT} and Leray-$\infty$, respectively, where the
forcing term on both models are rescaled appropriately to avoid
trivial dynamics.  The equations for the NS-$\infty$ and
Leray-$\infty$ are exactly the equations (\ref{nsa}) and
(\ref{leray-a}) together with the incompressibility condition except
that equations (\ref{rel-n}) and (\ref{rel-l}) are replaced by the
equations $\tilde{v} = -L^2\Delta \tilde{u}$ and $v = -L^2\Delta u$,
respectively.  Under the assumption that the scaling of the spectrum
as $\alpha \rightarrow \infty$ is identical to the scaling in the
range $k\alpha \gg 1$ for finite (small) $\alpha$ and sufficiently
long scaling ranges, we obtain a high resolution numerical calculation
of the sub-$\alpha$ scales. This assumption was verified in the finite
$\alpha$ calculation of the NS-$\alpha$ model for two-dimensions in
\cite{LKTT}.  We stress again here, that one has to rescale the
forcing appropriately in order to avoid trivial dynamics for large
values of $\aa$, and decaying turbulence at the limit when
$\aa\ra\infty$ \cite{IT}.

Let ${U}_k$ and $V_k$ denote the typical smoothed and unsmoothed
  velocities of an eddy of size ~$1/k$ for the Leray-$\alpha$ model.
  Similarly, let $\widetilde{U}_k$ and $\widetilde{V}_k$ denote the
  typical smoothed and unsmoothed velocities of an eddy of size~$1/k$
  for the NS-$\alpha$ model. Such `typical' velocities may be defined
  by the energy per unit area in the shell $[k,2k)$ as we will show in
  the next section. From our simulations of 2d NS-$\infty$ in
  \cite{LKTT}, the energy spectrum of the smoothed velocity $\tilde u$
  scales as $k^{-7}$. In this paper we will show from numerical simulations that the energy spectrum of the smoothed velocity
  $u$ of the 2d Leray-$\infty$ model scales as $k^{-5}$. These scalings can also be derived analytically (see \cite{LKTT} and section \ref{NsavsLeray} below) under the assumption that an eddy of size $1/k$, for $k\alpha\gg 1$, has a typical time scale comparable to the inverse of the square root of the enstrophy contained in this eddy.  That is, the dominant direct cascading quantity, which is the enstrophy in the 2d NS-$\alpha$ and Leray-$\alpha$, dictates these typical time scales.  Specifically, under this assumption, the governing time scales for an eddy of size $1/k$
  in each model are given by $(k \widetilde{V}_k)^{-1}$ (for NS-$\alpha$) and
  $(k\sqrt{{U}_k{V}_k}\;)^{-1}$ (for Leray-$\alpha$). We assert 
that the difference in the dominant forward cascading conserved
quantities in these two models is what leads to the different power
laws. Our numerical results in 2d for two different $\alpha$-models, with
different forward cascading conserved quantities, support this
assertion.

Based on our studies in 2d, we extrapolate our conclusions to
  the 3d case as follows. For the 3d NS-$\alpha$ model and 3d
  Leray-$\alpha$ model, the governing time scales for an eddy of
  size~$1/k$, for $k\alpha\gg 1$, must be given by $(k\sqrt{
    \widetilde{U}_k \widetilde{V}_k}\;)^{-1}$ and $(k
  {V}_k)^{-1}$, respectively.  This is because the energy conserved
  (in the absence of forcing and viscosity) in the 3d NS-$\alpha$ is
  given by $\int_\Om \tilde{u}\cdot\tilde{v}\;dx$ while in the 3d
  Leray-$\alpha$ it is given by $\int_\Om v\cdot v\;dx$.  Accordingly,
  we assert that for $k\alpha\gg 1$, the energy spectra of the
  smoothed velocity fields in the 3d case will scale as $k^{-11/3}$
  (steeper than $k^{-3}$ as originally suggested in \cite{FHTP}) for
  the 3d NS-$\alpha$, and as $k^{-17/3}$ (steeper than $k^{-13/3}$
  proposed in \cite{CHOT}) for the 3d Leray-$\alpha$ model.  This assertion is yet to be confirmed computationally in future work. Our prediction of $k^{-17/3}$ power law for the smoothed energy spectrum of the 3d Leray-$\alpha$ model corresponds to one of the three candidate power laws derived in \cite{CHOT}.  The idea that the average velocity of an eddy of size of the order $1/k$ can be evaluated in three different ways, which will then lead to three different power laws, was in fact first introduced in \cite{CHOT}.      

Throughout the paper we denote by $\tau_k$ the characteristic time
scale of an eddy of size $1/k$, $\Omega = [0,L]^2$. We denote the
rough and smoothed vorticities by $q=\nabla\times v$
(or~$\tilde{q}=\nabla\times \tilde{v}$) and $\omega=\nabla\times u$
(or $\tilde{\omega}=\nabla\times \tilde{u}$), respectively.  The paper
is organized as follows.  In section \ref{NsavsLeray} we derive the
power laws for the 2d Leray-$\alpha$ model and then give a comparison
to the corresponding power laws of the 2d NS-$\alpha$ equations.  In
section \ref{Results}, we give a brief review of the numerical results
in \cite{LKTT} and then present our numerical results for the 2d
Leray-$\alpha$.  In the last section, we give a summary of our main
results and give a brief description of how this study can help us
predict the unknown power laws for the 3d NS-$\alpha$ and 3d
Leray-$\alpha$ equations.  As mentioned above, our predictions, based
on this study, on the power laws for the two models just mentioned,
are different from those suggested in \cite{CHOT,FHTP}.  In those
works, the $k^{-3}$ and $k^{-13/3}$ power laws for the 3d NS-$\alpha$
and 3d Leray-$\alpha$ model, respectively, were proposed under the
assumption that the time scale which governs the small scales is
determined by the smoothed velocity field alone (even though there were two other candidate power laws derived in \cite{CHOT}).

We dedicate this work to our friend and colleague Darryl D. Holm on the
occasion of his 60th birthday in acknowledgement of his continuing
support and inspiration in stimulating scientific interactions and
discussions over the past years and many to come.

\section{Navier-Stokes-$\alpha$ vs. Leray-$\alpha$ model in two
  dimensions}\label{NsavsLeray}
In this section we give a comparison between the two $\alpha$-models.
For completeness we briefly present the analytical arguments for the
different power laws of the energy spectra which arise in the 2d
Leray-$\alpha$ equations.  For complete details we direct the reader
to look at the derivation of power laws of the 2d NS-$\alpha$ in
\cite{LKTT} (see also \cite{CHOT, CHTi,CLT,FOIAS,FHTP,FMRT,ILT} for
the analytical calculation of the power laws of energy spectra for the
other $\alpha$-models).  To compute the scaling of the smoothed energy
spectrum $E^u(k)$ of the 2d Leray-$\alpha$ in the wavenumber regime
$k\alpha\gg 1$ in the forward enstrophy inertial subrange, we start by
splitting the flow into the three wavenumber ranges $[1,\k),
[\k,2\k),[2\k,\infty)$.  For a wavenumber $\k$, we define the
component $u_\k$ of a velocity field $u$ by
\begin{equation}
u_k:=u_\k (x)= \sum_{|\xi|=\k}  \hat{u}(\xi)e^{i\frac{2\pi}{L}\xi\cdot x},
\end{equation}
and the component $u_{\k',\k''}$ with a range of wavenumbers $[\k',\k'')$ by
\begin{equation}
u_{\k',\k''}:=u_{\k',\k''}(x) = \sum_{\k'\leq\k<\k''}u_\k.
\end{equation} 
$E^u(k)$ is then the energy spectrum associated to 
$$\mathfrak{e}_k=\frac{1}{2}\ang{\|u_{k,2k}\|_{L^2(\Om)}^2}$$
which is the average (with respect to an infinite time average measure \cite{FMRT}) smoothed energy per unit mass of eddies of linear size $l\in (\frac{1}{2\k},\frac{1}{\k}]$.  

We assume $\k_f < \k$,
where $\k_f$ is the forcing wavenumber, since we are interested on the
effects of the Leray-$\alpha$ model in the enstrophy cascade regime. 
We 
decompose the $u$, $v$ and the $\nabla \times u$ and $\nabla \times v$
corresponding to the three wavenumber ranges.  We then write the enstrophy
balance equation for the Leray-$\alpha$ model for an eddy of size~
$\sim \k^{-1}$.  Taking an ensemble
average (with respect to infinite time average measure) of the enstrophy balance equation we get
\begin{equation}\label{TAET1}
\nu \ang{(\|\Delta u_{k,2k}\|_{L^2(\Om)}^2+\aa^2\|\nabla\Delta u_{k,2k}\|_{L^2(\Om)}^2)} = \ang{Z_\k} -\ang{Z_{2\k}},
\end{equation}
where $Z_\k$ may be interpreted as the net amount of enstrophy per unit time that is
transferred into wavenumbers larger than or equal to $\k$. Similarly, $Z_{2\k}$
represents the net amount of enstrophy per unit time that is transferred
into wavenumbers larger than or equal to $2\k$. Thus, $Z_\k -
Z_{2\k}$ represents the net amount of enstrophy per unit time that is
transferred into wavenumbers in the interval $[\k,2\k)$.
We  then
rewrite the averaged enstrophy transfer equation (\ref{TAET1}) as
$$\nu \k^5E_\alpha(\k) \sim \nu \int_\k^{2\k} \k^4E_\alpha(\k)d\k \sim
\ang{Z_\k}-\ang{Z_{2k}},$$
where $E_\alpha(k)$ is the energy spectrum associated to 
\begin{equation}
\mathfrak{e}_k^\aa = \frac{1}{2} \ang{\|u_{k,2k}\|_{L^2(\Om)}^2+\aa^2\|\nabla u_{k,2k}\|_{L^2(\Om)}^2},
\end{equation}
which is the average energy per unit mass of eddies of linear size 
$l\in (\frac{1}{2\k},\frac{1}{\k}]$. 

Thus, as long as $\nu \k^5E_\alpha(\k) \ll \ang{Z_\k}$ (that is,
$\ang{Z_{2\k}}\approx\ang{Z_\k}$, there is no leakage of enstrophy due to
dissipation), the wavenumber $\k$ belongs to the inertial range.  In the forward cascade inertial subrange, we follow Kraichnan \cite{K67} (see also \cite{FOIAS}) and
postulates that the eddies of size larger than $1/k$ transfer their energy to eddies of size smaller than $1/(2k)$ in the time $\tau_k$ it takes to travel their length $\sim 1/k$.  That is,
\begin{equation}
  \tau_\k \sim \frac{1}{\k \overline{U}_\k},
\label{tau}
\end{equation}
where $\overline{U}_\k$ is the average velocity of eddies of size $\sim 1/\k$. Since
there are two different velocities in the Leray-$\alpha$ model, there are three physically relevant possibilities for this average velocity, namely
\begin{eqnarray*}\label{3-vel}
  &&\hskip-.28in
  U_\k^{0}
  = \ang{ \frac{1}{L^2} \int_{\Om} |v_{\k,2k}(x)|^2
    dx}^{1/2} \sim \left(\int_\k^{2\k}(1+\alpha^2\k^2)E_\alpha(\k)dk\right)^{1/2}\sim \left( \k (1+\alpha^2 \k^2) E_{\alpha} (\k) \right)^{1/2},   \\
  &&\hskip-.28in
  U_\k^{1} = \ang{ \frac{1}{L^2} \int_{\Om} u_{\k,2k}(x) \cdot v_{\k,2k}(x)
    dx}^{1/2} \sim \left(\int_\k^{2\k}E_\alpha(\k)d\k\right)^{1/2}\sim \left( \k E_{\alpha} (\k) \right)^{1/2},   \\
  &&\hskip-.28in
  U_\k^{2} = \ang{ \frac{1}{L^2} \int_{\Om} |u_{\k,2k}(x)|^2
    dx}^{1/2} \sim \left(\int_\k^{2\k}\frac{E_\alpha(\k)}{(1+\alpha^2\k^2)}d\k\right)^{1/2} \sim \left( \frac{\k E_{\alpha} (\k) }{ 1+\alpha^2 \k^2}
  \right)^{1/2}.
\end{eqnarray*}
These define the aforementioned `typical' velocities, in particular 
$V_k = U^0_k$ and $U_k = U^2_k$. Corresponding
  definitions may be made for $\widetilde{V}_k$ and $\widetilde{U}_k$
  using the variables for the NS-$\alpha$ model, see \cite{LKTT}.
Thus,
\begin{equation}\label{n}
U_\k^n \sim \frac{(\k E_\alpha(\k))^{1/2}}{(1+\alpha^2\k^2)^{(n-1)/2}}, ~~ (n
= 0,1,2).
\end{equation}

We may therefore write the typical time scale of an eddy of size $\k^{-1}$ in (\ref{tau}) as
\begin{equation}
\tau_\k^n \sim \frac{1}{\k U_\k^n}
 = \frac{(1+\alpha^2\k^2)^{(n-1)/2}}{\k^{3/2}(E_\alpha(\k))^{1/2}}, ~~(n = 0, 1, 2).
\label{taun}
\end{equation}
That is, in keeping with the
  historical approach to the problem \cite{FOIAS,FHTM}, there are in principal
  three different time scales and it is left to empirical evidence to
  infer the correct time scale for a particular $\alpha$-model.  

The enstrophy dissipation rate $\eta_\alpha$ which is a constant
equal to the flux of enstrophy from wavenumber $\k$ to $2\k$ is given by
\begin{equation}
\eta_\alpha \sim \frac{1}{\tau_\k^{n}} \int_k^{2\k} \k^2 E_{\alpha} (\k) d\k \sim
\frac{\k^{9/2} \left( E_{\alpha} (\k) \right)^{3/2} }{(1+\alpha^2
\k^2)^{(n-1)/2}},
\end{equation}
and hence
\[
E_{\alpha}(\k) \sim \frac{\eta_\alpha^{2/3} (1+\alpha^2\k^2)^{(n-1)/3}}{\k^{3}}.
\]
Thus, the energy spectrum of the smoothed velocity $u$ is given by
\begin{eqnarray}\label{3-spec}
&&\hskip-.28in
E^u(\k)
\equiv
\frac{E_{\alpha} (\k)}{1+ \alpha^2 \k^2} \sim \left\{
\begin{array}{ll}   \displaystyle{
{\eta_{\alpha}^{2/3}}{\k^{-3}},}
\qquad & \mbox{when  }
\k\alpha \ll 1\,, \\
\displaystyle{\frac {\eta_{\alpha}^{2/3}}{\alpha^{2(4-n)/3}}
\k^{-(17-2n)/3}}, \qquad & \mbox{when  } \k\alpha \gg 1\,.
\end{array} \right.
\end{eqnarray}
Therefore, depending on the average velocity of an eddy of size
$\k^{-1}$ for the Leray-$\alpha$ model, we obtain three possible scalings
of the energy spectrum, $\k^{-(17-2n)/3}$, $(n = 0,1,2)$ all of which
decay steeper than the Kraichnan $\k^{-3}$ power law, in the subrange
$\k\alpha \gg 1$. {\it The goal, which we stress here again, is to infer the correct time scale by measuring the scaling exponent of energy spectra computed from high-resolution numerical simulation data.}

We summarize some points of comparison between the two models in Table
\ref{nsavsla}.  In the absence of viscosity $\nu$ and the forcing $f$,
the two conserved quantities, namely the energy and enstrophy, for the
two models, in two dimensions, are specified in the first block-row of
Table \ref{nsavsla}. Since the energy in 2d flow goes upscale
{(\cite{FOIAS, FMRT, K67, LKTT})}, we are more interested in the
enstrophy which has its dominate cascade downscale {(\cite{FOIAS,
    FMRT, K67, LKTT})}. For the NS-$\alpha$ model, the enstrophy
conserved is $\Omega_\alpha = \int_\Omega |\tilde{q}|^2 dx$, while for
the Leray-$\alpha$, the conserved enstrophy is given by
$\Omega_\alpha^L = \int_\Omega q\cdot \omega\;dx$.  It is for this
reason that the characteristic time scale for eddies of size smaller
than the length scale $\alpha$, for the two models, differ. The second block-row gives the three possible characteristic time scales, and the corresponding scaling predictions, three for each of the models.
The notation in Table 1 defaults to that for NS-$\alpha$; replace $\widetilde{U}_k \mbox{ and } \widetilde{V}_k$ by $U_k
  \mbox{ and }V_k$, respectively, in the formula for $\tau_k$ for Leray-$\alpha$.

These three possibilities for $\tau_k$ and the corresponding power
laws are given in Table~\ref{nsavsla}.  In \cite{LKTT}, it was shown
that the energy spectrum of the 2d NS-$\infty$ equations attains a
power law of $k^{-7}$ as the resolution is increased.  The convergence
of the spectral scaling is presented in the third block-row in Table
\ref{nsavsla}. A similar study for the 2d Leray-$\infty$ shows a
convergence to a power law of $k^{-5}$ as we shall show in the next
section.

\begin{table}[h]
\caption{Comparison between the NS-$\alpha$ and Leray-$\alpha$ in two dimensions (in the second block-row, $\widetilde{U}_k$ and $\widetilde{V}_k$ should be replaced by $U_k$ and $V_k$ respectively to obtain the $\tau_k$ for Leray-$\alpha).$}
\begin{indented}
\item []
  {\begin{tabular}{lll}\br   & NS-$\alpha$  & Leray-$\alpha$\\\\ 
     Ideal invariants: & Energy $e_\alpha = \int_\Omega \tilde{u}\cdot \tilde{v}\;dx$&Energy $e^L_\alpha = \int_\Omega u\cdot v\;dx$  \\
&Enstrophy $\Omega_\alpha = \int_\Omega |\tilde{q}|^2\;dx$ & Enstrophy $\Omega^L_\alpha = \int_\Omega q\cdot\omega\;dx$\\
\hline
Expected scaling \\
in the range $k\alpha\gg 1$ & & \\
\\
if $\tau_k = (k\widetilde{V}_k)^{-1}$&$k^{-7}$ &$k^{-17/3}$\\
if $\tau_k = \left(k\sqrt{\widetilde{U}_k \widetilde{V}_k}\right)^{-1}$&$k^{-19/3}$&$k^{-5}$\\
if $\tau_k = (k\widetilde{U}_k)^{-1}$ &$k^{-17/3}$ &$k^{-13/3}$\\
\hline
Convergence of $k^{-\gamma}$\\
as resolution is increased\\\\
$1024^2\quad \gamma=$&7.4 & 5.5\\
$2048^2\quad\gamma =$&7.1&5.2\\
$4096^2\quad\gamma=$&7.0&5.0\\
      \br
\end{tabular}}
\end{indented}
\label{nsavsla}
\end{table}
\section{Numerical results}\label{Results}
\subsection{Details of the numerical simulation}
The Leray-$\alpha$ equations were solved numerically in a periodic
domain of length $L=1$ on each side.  The wavenumbers $k$ are thus
integer multiples of $2\pi$.  A pseudospectral code was used with
fourth-order Runge-Kutta time-integration.  Simulations were carried
out with resolutions ranging from $1024^2$ up to $4096^2$ on the
Advanced Scientific Computing QSC machine at the Los Alamos National
Laboratory.  To maximize the enstrophy inertial subrange, the forcing
is applied in the wavenumber shells $2<k<4$.  We also add a
hypoviscous term $\mu(-\Delta)^{-1} v$ which provides a sink in the low
wavenumbers.  To discern a clear power-law of the Leray-$\alpha$ model
spectrum, we consider data from simulation of the Leray-$\infty$
equations
\begin{eqnarray}\label{leray-inf}
\partial_t v  -(u\cdot\nabla) v &=
-\nabla p +\nu\Delta v+f\\
\nabla\cdot u =\nabla\cdot v&=0\\
v = -L^2\Delta u,
\end{eqnarray}

Similar to the case of the 2d NS-$\infty$ equations \cite{IT, LKTT},
this allows us to see the scaling of the Leray-$\alpha$ model energy
spectrum without contamination by finite-$\alpha$ effects.

\begin{figure*}[ht]
\centering
\subfigure[]{
\label{fig: P11} 
\includegraphics[width=7cm]{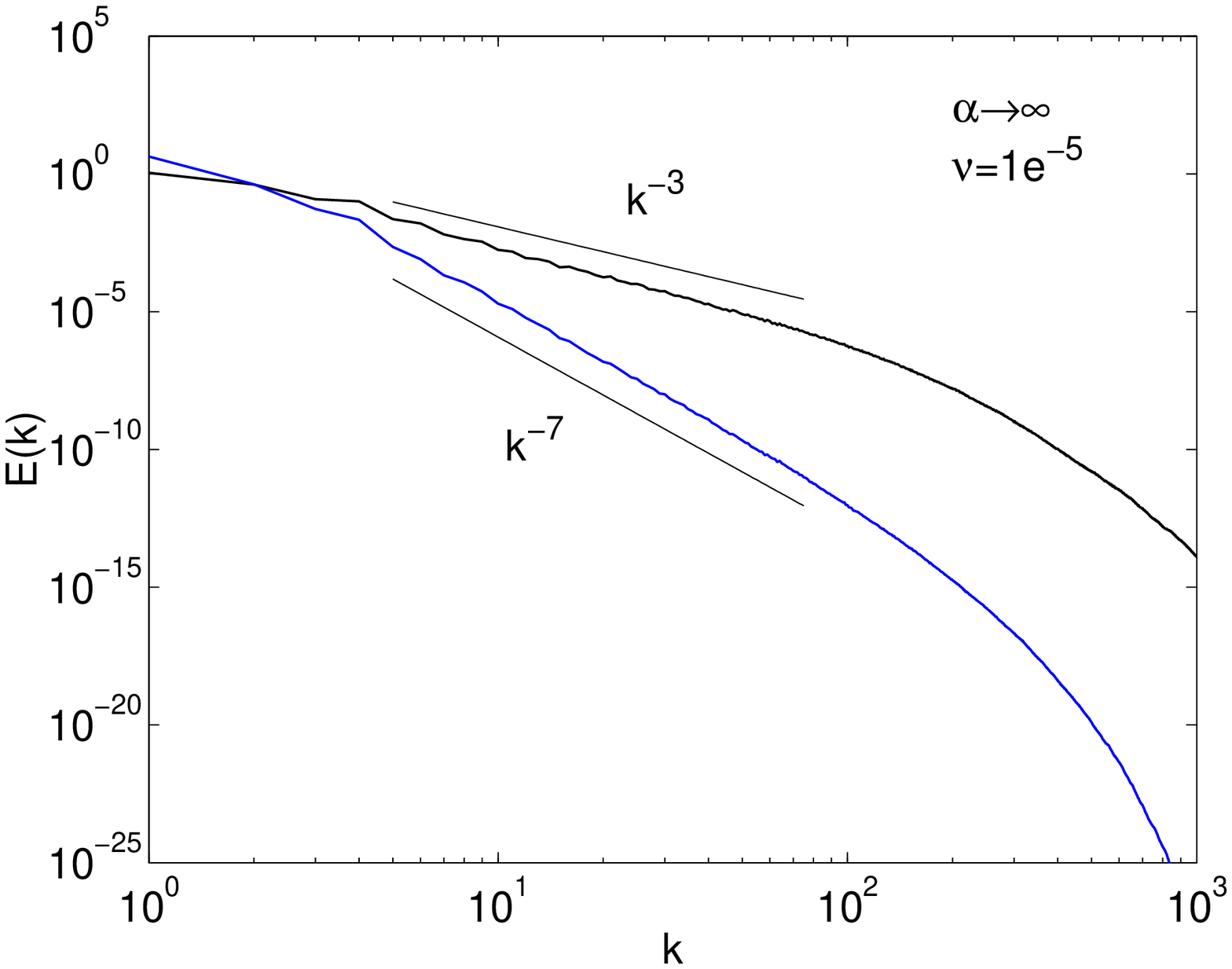}
}
\subfigure[]{%
\label{fig: P12}
\includegraphics[width=7cm]{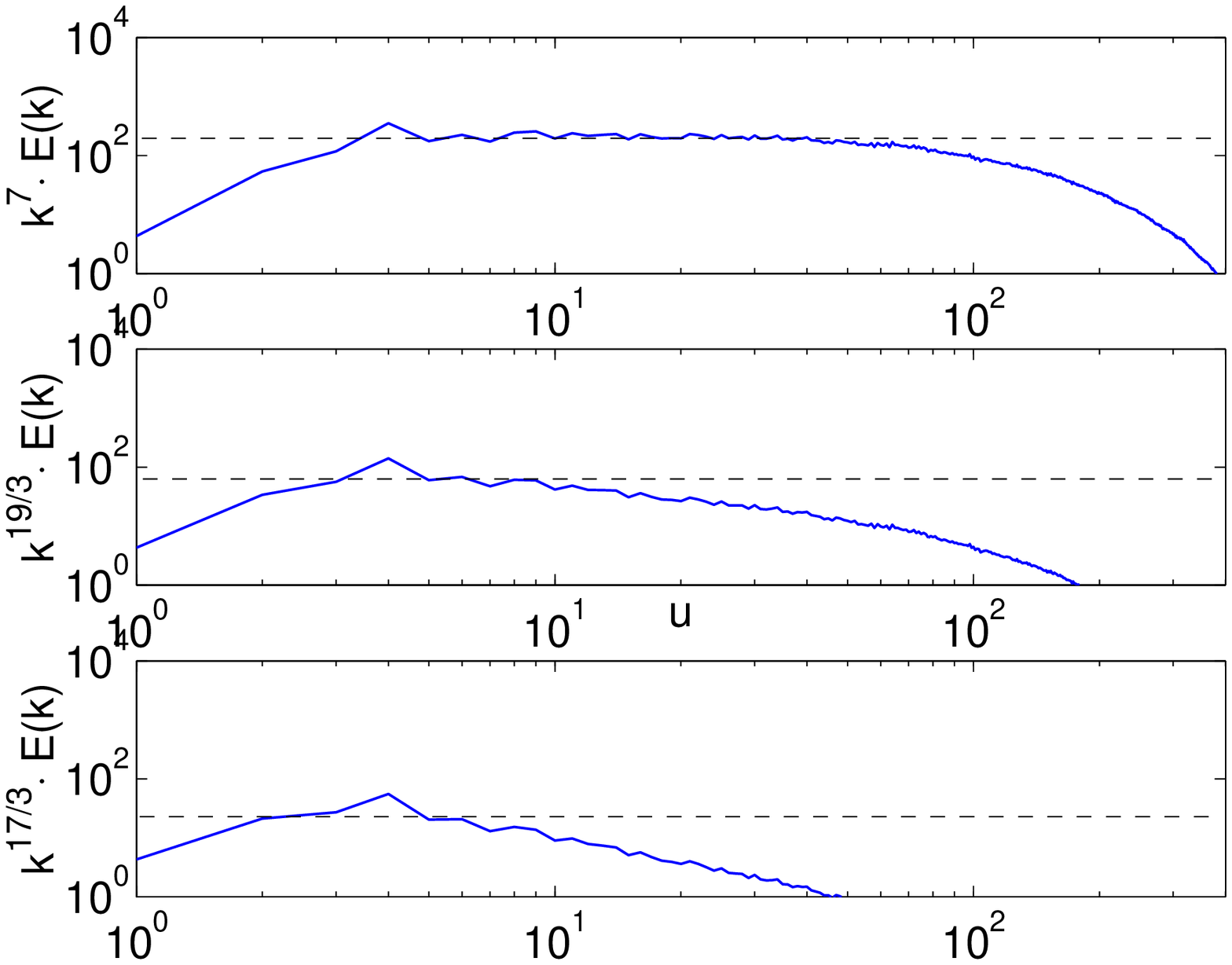}
}
\caption{{Energy spectra $E(k) = E^u(k)$ from a $4096^2$ simulation of
    the NS-$\alpha$ equations. (a): the blue line is the spectrum for
    NS-$\infty$, the black line for the NSE ($\alpha = 0$).  (b) top
    to bottom: the energy spectrum for NS-$\infty$ compensated
    by $k^{7}, k^{19/3}, \mbox{and } k^{17/3}$, respectively. The
    compensated spectrum in the top panel is flat in the range $6 < k
    < 40$, indicating the nominal range over which the $k^{-7}$
    scaling holds.}}
 \label{fig: Pn}
\subfigure[]{
\label{fig: P21} 
\includegraphics[width=7cm]{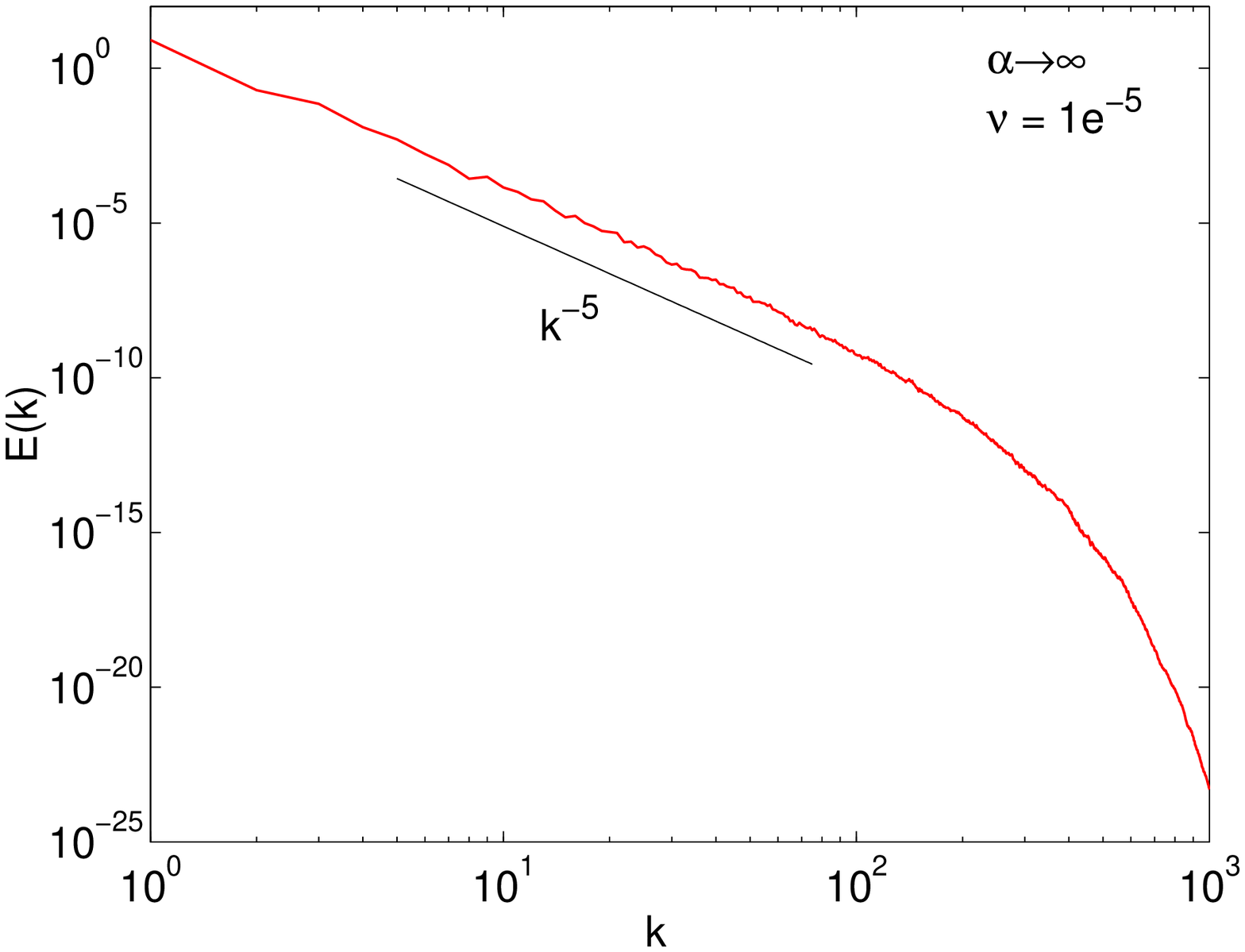}
}
\subfigure[]{%
\label{fig: P22}
\includegraphics[width=7cm]{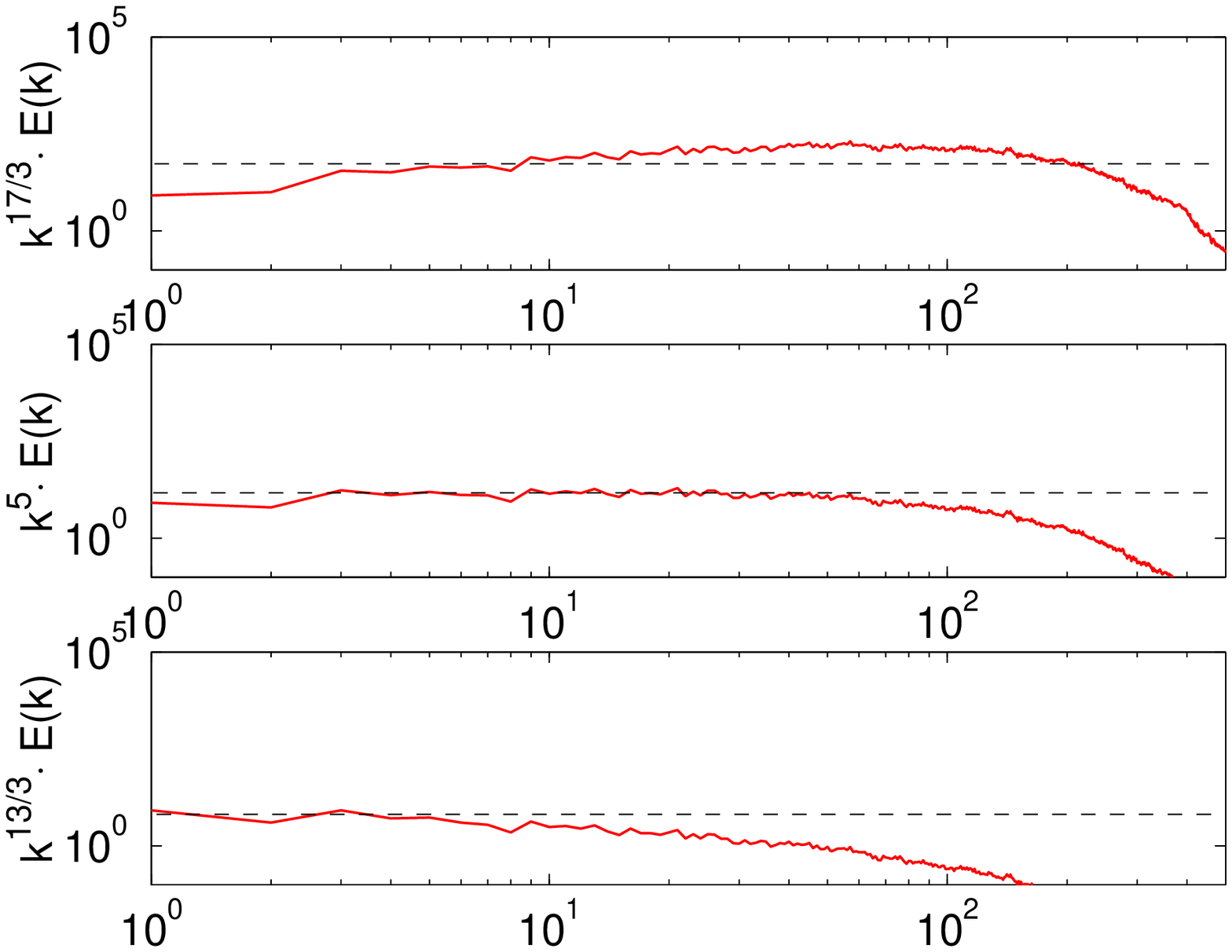}
}
\caption{{Energy spectrum $E(k) = E^u(k)$ from a $4096^2$
    simulation of the Leray-$\alpha$ equations. (a): energy spectrum
    for the Leray-$\infty$. (b) top to bottom: energy spectrum of
    Leray-$\alpha$ compensated by $k^{17/3}, k^5, \mbox{and }
    k^{13/3}$, respectively. The compensated spectrum in the middle
    panel is flat in the range $7 < k < 50$, indicating the nominal
    range over which the $k^{-5}$ scaling holds.}}
\label{fig: Pl}
\end{figure*}
\subsection{Results for the Leray-$\infty$ equations} 
In Figures~\ref{fig: Pn} and \ref{fig: Pl} we
  use the notation $E(k)$ for $E^u(k)$. In Figure \ref{fig: Pn} we present the main numerical results from \cite{LKTT}
for the 2d NS-$\infty$ equations, showing the $k^{-7}$ scaling of
the energy spectrum.  In Figure~\ref{fig: Pl} we show that the
2d Leray-$\infty$ energy spectrum attains a $k^{-5}$ power law in a
$4096^2$~resolution simulation.  From Table~\ref{nsavsla}, the
  scaling $k^{-5}$ stems, based on the analytical arguments in
  section~\ref{NsavsLeray}, from a characteristic time scale given by
  $\tau_k = \left(k\sqrt{{V}_k{U}_k}\right)^{-1}$, which is comparable to the inverse of the square root of the enstrophy of an eddy of the size $1/k$.  To see this, recall from section \ref{NsavsLeray} that the typical smoothed and unsmoothed velocity
  of eddy of size~$\sim 1/k$ are given  by
\begin{eqnarray}
{U}_k &= \frac{1}{L^2}\ang{\|u_{k,2k}(x)\|_{L^2(\Omega)}},\\
{V}_k &= \frac{1}{L^2}\ang{\|v_{k,2k}(x)\|_{L^2(\Omega)}},
\end{eqnarray}
then, we can define
\begin{eqnarray}
\mathcal{W}_k &= \frac{1}{L^2}\ang{\|\nabla\times u_{k,2k}\|_{L^2(\Om)}},\\
Q_k &= \frac{1}{L^2}\ang{\|\nabla\times v_{k,2k}\|_{L^2(\Om)}},
\end{eqnarray}
as the smoothed and unsmoothed enstrophy per unit area in the shell
$[k,2k).$ Now, observe that
$$\tau_k = \left(k\sqrt{{V}_k{U}_k}\right)^{-1}\sim \left(\sqrt{{Q}_k{\mathcal{W}}_k}\right)^{-1}.$$
 Therefore, the numerical results in Figure \ref{fig: Pl} clearly support our
  claim that the characteristic time scale $\tau_k
    =\left(k\sqrt{{V}_k{U}_k}\right)^{-1}$ determined by the
  dominant cascading quantity, namely the enstrophy $\int_\Om q\cdot
  \omega\; dx$ for the 2d Leray-$\infty$ model, governs the dynamics
  of eddies in the subrange $k\alpha\gg 1$.  This conclusion is
  consistent with our original prediction in \cite{LKTT} that, in
  general, the dominant cascading conserved quantity in the
  $\alpha$-model dictates the time scale associated with eddy of size $1/k$ and hence the power law of the energy spectrum
  $E^u(k)$ in the subrange $k\alpha\gg 1$.

\section{Conclusion}
The main goal of this study is to verify our claim in \cite{LKTT}
about the choice of particular characteristic time scales of eddy of size $1/k$, for $k\alpha\gg 1$, for
particular $\alpha$-model equations. Our results in \cite{LKTT} led us to
conclude that the choice depends on the form of the cascading
conserved enstrophy, which is the dominant forward cascading quantity.  To verify this conclusion, we perform a high
resolution simulation of the 2d Leray-$\alpha$ equations in the limit
as $\alpha\ra\infty$ similar to our study of the 2d NS-$\infty$
equations.  We summarize the three steps to this study which verifies
these claims:
\begin{enumerate}
\item Identify the conserved quantities (in the absence of viscosity
  and forcing) for 2d Leray-$\alpha$, and the dominant one in the forward cascade regime.
\item Calculate the power laws using semi-rigorous arguments as in
  \cite{CHTi, CLT, FOIAS, FHTP, FMRT, ILT, LKTT}.  This will give us the
  three possible power laws of the energy spectrum in the wavenumber regime $k\alpha\gg 1$.
\item Perform a high-resolution simulation to identify which one of
  the power laws of the energy spectrum calculated in step (ii) actually arise for the 2d Leray-$\alpha$ model.

  \vspace*{4mm} As we have speculated in \cite{LKTT}, the scaling
  exponent in the wavenumber regime $k\alpha\gg 1$ will be governed by
  the time scale of the dominant cascading conserved quantity in that
  regime.  If we extend the same argument to predict the scaling for
  the 3d NS-$\alpha$ and the 3d Leray-$\alpha$ model equations then we
  obtain the following predictions.  Since $E^{uv} =
  \frac{1}{2}\int_{[0,L]^3}\tilde{u}\cdot \tilde{v}\;dx$ and $E^{vv} =
  \frac{1}{2}\int_{[0,L]^3}v\cdot v\;dx$ are the conserved energy
  which are the dominant cascading quantities for the 3d NS-$\alpha$
  and 3d Leray-$\alpha$, respectively, then we predict the scaling of
  $E^u(k)\sim k^{-11/3}$ for the 3d NS-$\alpha$ (that is, steeper than
  the $k^{-3}$ proposed in \cite{FHTP}) and the scaling of
  $E^u(k)\sim k^{-17/3}$ for the 3d Leray-$\alpha$ (that is, steeper
  than the $k^{-13/3}$ suggested in \cite{CHOT} in the wavenumber
  regime $k\alpha\gg 1$. Our prediction of $k^{-17/3}$ power law corresponds to one of the power laws initially derived in \cite{CHOT} as candidate power law for the smoothed energy spectrum of the 3d Leray-$\alpha$ model. The verification of these possibilities in the 
3d case will be explored in future work.

\end{enumerate}  
\section{Acknowledgments}
We are grateful to Mark A. Taylor for his continued collaboration on
this study.  E.~Lunasin was supported by UC Irvine and the NSF grant no. DMS-0504619
when this work was initiated. S. Kurien was supported by the NNSA of the
U.S. DOE at Los Alamos National Laboratory under contract no.
DE-AC52-06NA25396, partially supported by the Laboratory Directed
Research and Development program and the DOE Office of Science
Advanced Scientific Computing Research (ASCR) Program in Applied
Mathematics Research. The work of E. S. Titi was supported in part by
the NSF grants no. DMS-0504619 and no. DMS-0708832, and the
ISF grant no. 120/6.

\end{document}